\documentstyle[12pt]{article}

\begin{document}

\begin{center}
{\large {\bf Scalar and spinor solutions in the spacetimes of vacuumless
defects} }
\end{center}

\centerline{Sandro G. Fernandes $^{a}$\footnote{sandrofernandes@funesa.al.gov.br}, Geusa de A. Marques$^{b}$\footnote{gmarques@df.ufcg.edu.br} 
and V. B. Bezerra $^{c}$\footnote{valdir@fisica.ufpb.br}}

\begin{center}
$^a$ Departamento de Matem\'atica, Universidade Estadual de
Alagoas,\\[0pt]
57601-460 Palmeira dos Indios, Al, Brazil.

$^b$ Departamento de F\'{\i}sica, Universidade Federal de Campina Grande,\\[%
0pt]
58109-790 Campina Grande, Pb, Brazil.

$^c$ Departamento de F\'{\i}sica, Universidade Federal da Para\'{\i}ba,\\[0pt%
]
58051-970 Jo\~ao Pessoa, Pb, Brazil.
\end{center}

\vskip 0.6cm

\centerline{Abstract} \noindent 

\hskip0.7cm We obtain solutions of the Klein-Gordon and Dirac equations in
the gravitational fields of vacuumless defects. We calculate the energy
levels and the current, respectively, in the scalar and spinor cases. In all
these situations we emphasize the role played by the defects on the
solutions, energy and current.

\section{Introduction}

The study concerning the interaction of a quantum mechanical system with a
gravitational field has been an exiting research field\cite{P1}-\cite{Chan}.
One of the conclusions of these investigations is that the energy levels of
an atom placed in a gravitational field are shifted as a result of the its
interaction with the spacetime curvature(local aspects)\cite{P2} as well as
with the spacetime topology (global aspects)\cite{russo}. Therefore, it is
also important to investigate the role played by the topological features of
a given gravitational field interacting with a quantum system. More
recently, other investigations on this subject have been carried out \cite%
{russo,Sak} motivated by the fact that these can have considerable as well
as possible observational interest.

One of the fascinating aspects of the gauge theories is that they can create
topological objects called topological defects, that appear due to the
spontaneous symmetry breaking. Among these defects we can have domain walls,
cosmic strings, global monopoles\cite{P14} and their hybrid\cite{P14a}.
These defects have been studied in models where the scalar potential , $%
V(\Phi )$, that appears in the Lagrangian that describes the system,
possesses an absolute minimum. However, topological defects can also be
formed when the potential $V(\Phi )$ does not possess a minimum, but, is a
monotonically decreasing function of the scalar field. These defects are
called vacuumless defects\cite{Cho}.

The aim of this paper is to find the solution of the Klein-Gordon equation
and study the energy shifts associated with scalar particles in the
spacetimes generated by a gauge vacuumless string, a global vacuumless
string and a vacuumless global monopole. In the spinor case, we solve the
Dirac equation and find the currents associated with relativistic particles
in these backgrounds.

In order to do these investigations, we will consider a scalar and spinor
quantum particles embedded in a classical background gravitational field. In
the scalar case, the behavior of the particle is described by the covariant
Klein-Gordon equation

\begin{equation}
\left[ \frac{1}{\sqrt{-g}}\partial _{\alpha }(\sqrt{-g}g^{\alpha \beta
}\partial _{\beta })-\mu ^{2}\right] \Psi =0,  \label{eq2}
\end{equation}%
where ${\mu }$ is the mass of the particle and $\hbar =c=1$ units are chosen.

In the case of a spinor particle, we consider the generally covariant form
of the Dirac equation which is given by

\begin{equation}
\left[ i\gamma ^{\mu }\left( x\right) \left( \partial _{\mu }+\Gamma _{\mu
}\left( x\right) \right) -\mu \right] \Psi \left( x\right) =0,  \label{103}
\end{equation}%
where $\Gamma _{\mu }\left( x\right) $ are the spinor affine connections
which can be expressed in terms of the set of tetrad fields $e_{(a)}^{\mu
}(x)$ and the standard flat spacetime $\gamma ^{(a)}$ Dirac matrices as

\begin{eqnarray}
\Gamma_{\mu}=\frac{1}{4}\gamma^{(a)}\gamma^{(b)}e_{(a)}^{\nu}
(\partial_{\mu}e_{(b)\nu} - \Gamma^{\sigma}_{\mu \nu}e_{(b)\sigma}).
\label{fim}
\end{eqnarray}

The generalized Dirac matrices $\gamma ^{\mu }\left( x\right) $ satisfies
the anticommutation relations 
\[
\left\{ \gamma ^{\mu }\left( x\right) ,\gamma ^{\nu }\left( x\right)
\right\} =2g^{\mu \nu }\left( x\right) , 
\]
and are defined by 
\begin{equation}
\gamma ^{\mu }\left( x\right) =e_{\left( a\right) }^{\mu }\left( x\right)
\gamma ^{\left( a\right) },  \label{104}
\end{equation}
where $e_{\left( a\right) }^{\mu }\left( x\right) $ obeys the relation $\eta
^{ab}e_{\left( a\right) }^{\mu }\left( x\right) e_{\left( b\right) }^{\nu
}\left( x\right) =g^{\mu \nu };$ $\mu ,\,\nu =0,1,2,3$ are tensor indices
and $a,\,b=0,1,2,3$ are tetrad indices.

This paper is organized as follows. In section II, we solve the Klein-Gordon
and Dirac equations, determine the energy shifts and calculate the current
associated with relativistic particles in the spacetime of a gauge
vacuumless string. In section III, we perform similar calculations in the
background spacetime generated by a global vacuumless string. In section IV,
once again, we do similar calculations, but now , in the spacetime of a
vacuumless global monopole. Finally, in section V, we end up with ours
conclusions.

\section{Scalar and spinor particles in the spacetime of a gauge vacuumless
string}

\hskip0.7cm Firstly, let us consider the metric corresponding to a gauge
vacuumless string which is given by\cite{Cho}

\begin{equation}
ds^{2}=\left( 1+2\Phi \right) (-dt^{2}+dz^{2})+d\rho ^{2}+(1-4\Phi )\rho
^{2}d\varphi ^{2},  \label{5.11}
\end{equation}%
where 
\[
\Phi (\rho )=2\pi a^{2}\ln \left( \frac{\rho }{\delta }\right) , 
\]%
is the Newtonian potential, in which $\delta =\lambda ^{1/2}M^{-1}$ is the
size of the defect core and $\lambda ,M$ and $a$ are positive constants.

Initially, we will consider a scalar quantum particle in the spacetime
generated by a gauge vacuumless string, whose line element is given by eq. (%
\ref{5.11}). In this spacetime, the covariant Klein-Gordon equation turns
into 
\begin{equation}
\left\{ \left( 1-2\Phi \right) \left( \frac{\partial ^{2}}{\partial z^{2}}-%
\frac{\partial ^{2}}{\partial t^{2}}\right) +\frac{1}{\rho }\frac{\partial }{%
\partial \rho }+\frac{\partial ^{2}}{\partial \rho ^{2}}+\frac{1+4\Phi }{%
\rho ^{2}}\frac{\partial ^{2}}{\partial \varphi ^{2}}-\mu ^{2}\right\} \Psi
=0.\qquad \qquad  \label{5.16}
\end{equation}

The solution of eq.(\ref{5.16}) can be obtained writing $\Psi (t,\rho
,\varphi ,z)$ in the following form 
\begin{equation}
\Psi (t,\rho ,\varphi ,z)=R(\rho )e^{-iEt+im\varphi +ikz}  \label{5.17}
\end{equation}%
where $E$ , $m$ and $k$ are constants.

Substituting eq.(\ref{5.17}) into (\ref{5.16}), we get the equation

\begin{equation}
\frac{d^{2}R}{d\rho ^{2}}+\frac{1}{\rho }\frac{dR}{d\rho }-\left[ \frac{m^{2}%
}{\rho ^{2}}\left( 1+4\Phi \right) +\left( 1-2\Phi \right) \left(
E^{2}-k^{2}\right) -\mu ^{2}\right] R=0.  \label{5.18a}
\end{equation}

The solution of eq. (\ref{5.18a}) can be obtained if we assume that the
particle is restricted to move in a cylindrical narrow region of width $\rho
_{2}-\rho _{1}$, with $\rho _{2}\cong \rho _{1}>0$. This implies that the
potential $\Phi (\rho )$ is approximately constant and equal to $\Phi _{0}$.
With this hypothesis, the solution of (\ref{5.18a}) is

\begin{equation}
R(\rho )=C_{1}J_{\left\vert \nu \right\vert }(\widetilde{k}\rho
)+C_{2}N_{\left\vert \nu \right\vert }(\widetilde{k}\rho ),  \label{19}
\end{equation}%
where $J_{\left\vert \nu \right\vert }(\widetilde{k}\rho )$ and $%
N_{\left\vert \nu \right\vert }(\widetilde{k}\rho )$ are Bessel functions of
first and second kind, respectively, with $\widetilde{k}^{2}\equiv \left(
E^{2}-k^{2}\right) \left( 1-2\Phi _{0}\right) -\mu ^{2}$ and $\nu
^{2}=\left( 1-4\Phi _{0}\right) m^{2}$. If we consider that the boundaries $%
\rho =\rho _{1}$ $>0$ and $\rho =\rho _{2}>\rho _{1}$ are impenetrable, we
can demand that $\Psi (t,\rho ,\varphi ,z)$ vanishes on these boundaries and
outside the cylindrical region. In this case, we can determine the energy
spectrum of the particle, from the boundary conditions 
\begin{equation}
R(\rho _{1})=R(\rho _{2})=0.  \label{5.19}
\end{equation}

The condition expressed in eq.(\ref{5.19}) furnishes us the following
equation

\begin{equation}
\qquad \qquad J_{\left\vert \nu \right\vert }(\widetilde{k}\rho
_{1})N_{\left\vert \nu \right\vert }(\widetilde{k}\rho _{2})\bigskip
-J_{\left\vert \nu \right\vert }(\widetilde{k}\rho _{2})N_{\left\vert \nu
\right\vert }(\widetilde{k}\rho _{1})=0.  \label{5.20}
\end{equation}

In order to obtain the energy spectrum explicitly, we will consider the
situation in the which $\widetilde{k}\rho _{1}>>1$ and $\widetilde{k}\rho
_{2}>>1$. With these conditions, we can use the Hankel$^{\prime }$s
asymptotic expansions\cite{abramo} to obtain the following results

\begin{equation}
J_{\left\vert \nu \right\vert }(\widetilde{k}\rho _{1})\sim \sqrt{\frac{2}{%
\pi \widetilde{k}\rho _{1}}}\left[ \cos \left( \widetilde{k}\rho _{1}-\frac{%
\nu }{2}\pi -\frac{\pi }{4}\right) -\frac{4\nu ^{2}-1}{8\widetilde{k}\rho
_{1}}\sin \left( \widetilde{k}\rho _{1}-\frac{\nu }{2}\pi -\frac{\pi }{4}%
\right) \right]  \label{5.21}
\end{equation}%
and 
\begin{equation}
N_{\left\vert \nu \right\vert }(\widetilde{k}\rho _{1})\sim \sqrt{\frac{2}{%
\pi \widetilde{k}r_{1}}}\left[ \sin \left( \widetilde{k}\rho _{1}-\frac{\nu 
}{2}\pi -\frac{\pi }{4}\right) +\frac{4\nu ^{2}-1}{8\widetilde{k}\rho _{1}}%
\cos \left( \widetilde{k}a-\frac{\nu }{2}\pi -\frac{\pi }{4}\right) \right]
\label{casa}
\end{equation}%
and similarly for $J_{\left\vert \nu \right\vert }(\widetilde{k}\rho _{2})$
and $N_{\left\vert \nu \right\vert }(\widetilde{k}\rho _{2})$ , by
interchanging $\rho _{1}$ by $\rho _{2}$ .

Substituting eqs.(\ref{5.21}) and (\ref{casa}) and its similar equations for 
$J_{\left\vert \nu \right\vert }(\widetilde{k}\rho _{2})$ and $N_{\left\vert
\nu \right\vert }(\widetilde{k}\rho _{2})$ into eq.(\ref{5.20}) , we obtain
the following result 
\begin{equation}
\widetilde{k}^{2}\sim \left( \frac{\pi n}{\rho _{2}-\rho _{1}}\right) ^{2}+%
\frac{4\nu ^{2}}{\rho _{1}\rho _{2}}-\frac{1}{4\rho _{1}\rho _{2}}.
\label{5.23}
\end{equation}

Thus, remembering that $\widetilde{k}^{2}\equiv \left( E^{2}-k^{2}\right)
\left( 1-2\Phi _{0}\right) -\mu ^{2}$ , finally, we obtain from (\ref{5.23}%
), that the energy spectrum is given by

\begin{equation}
E=\sqrt{k^{2}+\mu ^{2}(1+2\Phi _{0})+\frac{4\left( 1-2\Phi _{0}\right) m^{2}%
}{\rho _{1}^{2}}-\frac{(1+2\Phi _{0})}{4\rho _{1}^{2}}}.  \label{5.25}
\end{equation}

Note that the energy depends upon the value $\Phi _{0}$ of the Newtonian
potential in the narrow cylindrical region.

Now, we will consider a spinor particle in this spacetime. To do this let us
solve the Dirac equation

\begin{eqnarray}
&&i\left[ \left( 1-\Phi _{0}\right) \gamma ^{(0)}\frac{\partial }{\partial t}%
+\gamma ^{(\rho )}\left( \frac{\partial }{\partial \rho }-\frac{\Phi _{0}}{%
\rho }\right) \right.  \nonumber \\
&&\qquad \qquad \left. +\frac{1+2\Phi _{0}}{\rho }\gamma ^{(\varphi )}\frac{%
\partial }{\partial \varphi }+\left( 1-\Phi _{0}\right) \gamma ^{(z)}\frac{%
\partial }{\partial z}\right] \Psi -\mu \Psi \left. =0\right.
\end{eqnarray}%
where $\gamma ^{(\rho )}=\gamma ^{(\rho )}\cos \varphi +\gamma ^{(\varphi
)}\sin \varphi $ and $\gamma ^{(\varphi )}=-\gamma ^{(\rho )}\sin \varphi
+\gamma ^{(\varphi )}\cos \varphi $ , $\Phi _{0}$ is the Newtonian potential
in the narrow cylindrical region defined in the scalar case and we have used
the following set of tetrads

\[
\begin{array}{ll}
e_{(0)}^{\mu } & =\left( 1+\Phi _{0}\right) \delta _{(0)}^{\mu },\vspace{0.3cm%
} \\ 
e_{(1)}^{\mu } & =\cos \varphi \delta _{(1)}^{\mu }-\frac{\left( 1+2\Phi
_{0}\right) }{\rho }\sin \varphi \delta _{(2)}^{\mu },\vspace{0.3cm} \\ 
e_{(2)}^{\mu } & =\sin \varphi \delta _{(1)}^{\mu }+\frac{\left( 1+2\Phi
_{0}\right) }{\rho }\cos \varphi \delta _{(2)}^{\mu },\vspace{0.3cm} \\ 
e_{(3)}^{\mu } & =\left( 1+\Phi _{0}\right) \delta _{(3)}^{\mu }.%
\end{array}%
\]

\bigskip

In order to take into account the symmetry let us adopt the following
representation for the Dirac matrices 
\[
\gamma ^{(0)}=\left[ 
\begin{array}{ll}
\sigma ^{3} & 0 \\ 
0 & -\sigma ^{3}%
\end{array}%
\right]; \quad \gamma ^{(2)}=\left[ 
\begin{array}{ll}
i\sigma ^{2} & 0 \\ 
0 & -i\sigma ^{2}%
\end{array}%
\right]; 
\]%
\begin{equation}
\gamma ^{(1)}=\left[ 
\begin{array}{ll}
-i\sigma ^{1} & 0 \\ 
0 & i\sigma ^{1}%
\end{array}%
\right];\quad \gamma ^{(3)}=\left[ 
\begin{array}{ll}
0 & 1 \\ 
-1 & 0%
\end{array}%
\right] .
\end{equation}%
where $\sigma ^{1}$ , $\sigma ^{2}$ and $\sigma ^{3}$ are the Pauli matrices.

\bigskip Let us write the spinor $\Psi $ in terms of the bi-spinors $\Psi
_{A}$ and $\Psi _{B}$, in the form $\Psi =\left( 
\begin{array}{l}
\Psi _{A} \\ 
\Psi _{B}%
\end{array}%
\right) $ and choose the following Ansatz 
\begin{equation}
\Psi _{A}=\Psi _{B}=\left[ 
\begin{array}{l}
\left( \sqrt{E+\mu }\right) u_{1}(\rho ) \\ 
\left( i\sqrt{E-\mu }\right) u_{2}(\rho )%
\end{array}%
\right] e^{-iEt+im\varphi +ikz}.
\end{equation}

\bigskip

Thus, we find the following equations for $u_{1}(\rho )$ and $u_{2}(\rho )$ 
\begin{equation}
\left\{ \frac{d^{2}}{d\rho ^{2}}+\frac{1}{\rho }\frac{d}{d\rho }-\frac{1}{%
\rho ^{2}}\left[ \left( m+\frac{1}{2}\right) (1-2\Phi _{0})-\frac{1}{2}%
\right] ^{2}+(E^{2}-\mu ^{2})\right\} u_{1}(\rho )=0  \label{5.42}
\end{equation}%
and 
\begin{equation}
\left\{ \frac{d^{2}}{d\rho ^{2}}+\frac{1}{\rho }\frac{d}{d\rho }-\frac{1}{%
\rho ^{2}}\left[ \left( m+\frac{1}{2}\right) (1-2\Phi _{0})+\frac{1}{2}%
\right] ^{2}+(E^{2}-\mu ^{2})\right\} u_{2}(\rho )=0,  \label{5.43}
\end{equation}
whose solutions are given in the general form by 
\begin{equation}
u_{i}(\rho )=C_{i,m}^{(1)}J_{\left\vert v+(i-1)\right\vert }(\varsigma \rho
)+C_{i,m}^{(2)}N_{\left\vert v+(i-1)\right\vert }(\varsigma \rho ),
\end{equation}%
where $i=1,2$ , $\varsigma ^{2}=E^{2}-\mu ^{2}$ , $v_{i}=\left( m+\frac{1}{2}%
\right) (1-2\Phi _{0})-\frac{1}{2}$ \ and $C_{i,m}^{(1)}$ and $C_{i,m}^{(2)}$
are constant bi-spinors.

Now, let us calculate the current, $j^{\mu }$, which can be written as

\begin{eqnarray}
j^{\mu } &=&\frac{1}{2\mu }\partial _{\lambda }\left( \overline{\Psi }\sigma
^{\mu \lambda }\Psi \right) +\frac{i}{4\mu }g^{\mu \lambda }\overline{\Psi }%
\stackrel{\leftrightarrow}{\partial}_{\lambda }\Psi +\frac{i}{4\mu }\overline{\Psi 
}\left( \left[ \partial _{\lambda }\gamma ^{\mu },\gamma ^{\mu }\right] +%
\left[ \gamma ^{\mu },\partial _{\lambda }\gamma ^{\mu }\right] \right) \Psi
\nonumber \\
&&+\frac{i}{2\mu }\overline{\Psi }\left[ \gamma ^{\lambda }\Gamma _{\lambda
},\gamma ^{\mu }\right] \Psi ,  \label{5.46}
\end{eqnarray}
where $\ \sigma ^{\mu \lambda }=\frac{i}{2}\left[ \gamma ^{\mu },\gamma
^{\lambda }\right] .$

In the spacetime of a gauge vacuumless string, we have the following results 
\begin{equation}
\begin{array}{ll}
\left[ \partial _{\lambda }\gamma ^{\lambda },\gamma ^{0}\right] & =\frac{2}{%
\rho }\left( 1-2\Phi _{0}\right) \gamma ^{(0)}\gamma ^{(\rho)}\bigskip \\ 
\left[ \gamma ^{\lambda }\Gamma _{\lambda },\gamma ^{0}\right] & =\frac{%
2\Phi _{0}}{\rho }\gamma ^{(0)}\gamma ^{(\rho)}\bigskip \\ 
\left[ \partial _{\lambda }\gamma ^{\lambda },\gamma ^{\varphi }\right] & =%
\frac{1}{\rho ^{2}}\left( 1-2\Phi _{0}\right) \left[ \gamma ^{(\varphi
)},\gamma ^{(\rho )}\right] \bigskip \\ 
\left[ \gamma ^{\lambda },\partial _{\lambda }\gamma ^{\varphi }\right] & =2 
\left[ \gamma ^{\lambda }\Gamma _{\lambda },\gamma ^{\varphi }\right] =\frac{%
2\Phi _{0}}{\rho ^{2}}\left[ \gamma ^{(\rho )},\gamma ^{(\varphi)}\right]
\bigskip \\ 
\left[ \partial _{\lambda }\gamma ^{\lambda },\gamma ^{z}\right] & =\frac{1}{%
\rho }(1-3\Phi _{0})\left[ \gamma ^{(z)},\gamma ^{(\rho )}\right] \bigskip
\\ 
\left[ \gamma ^{\lambda }\Gamma _{\lambda },\gamma ^{z}\right] & =\frac{\Phi
_{0}}{\rho }\left[ \gamma ^{(\rho )},\gamma ^{(z)}\right] \bigskip%
\end{array}
\label{ca}
\end{equation}%
and 
\begin{equation}
\begin{array}{ll}
\sigma ^{01} & =i\gamma ^{(0)}\gamma ^{(\rho )}\bigskip \\ 
\sigma ^{02} & =\frac{i}{\rho }(1+\Phi _{0})\gamma ^{(0)}\gamma ^{(\varphi
)}\bigskip \\ 
\sigma ^{03} & =i(1-\Phi _{0})\gamma ^{(0)}\gamma ^{(z)}\bigskip \\ 
\sigma ^{12} & =\frac{i}{2\rho }(1+2\Phi _{0})\left[ \gamma ^{(\rho
)},\gamma ^{(\varphi )}\right] \bigskip \\ 
\sigma ^{13} & =\frac{i}{2}(1-\Phi _{0})\left[ \gamma ^{(\rho )},\gamma
^{(z)}\right] \bigskip \\ 
\sigma ^{23} & =\frac{i}{2\rho }(1+\Phi _{0})\left[ \gamma ^{(\varphi
)},\gamma ^{(z)}\right].
\end{array}
\label{ca1}
\end{equation}

Therefore, using the expression for the current given by the eq.(\ref{5.46})
and the results given in (\ref{ca}) and (\ref{ca1}), we obtain the following
results for the components of the current

\begin{eqnarray}
j_{t}&=&\vec{\nabla}_{\Phi _{0}}.\vec{P}+j_{(t)}{}_{conv.}\\
j_{\rho }&=&-\partial _{t}P_{\rho }+\left( \vec{\nabla}_{\Phi _{0}}\times \vec{%
M}\right) _{\rho }+j_{(\rho )}{}_{conv.}\\
j_{\varphi }&=&-\partial _{t}P_{\varphi }+\left( \vec{\nabla}_{\Phi
_{0}}\times \vec{M}\right) _{\varphi }+j_{(\varphi )}{}_{conv.}\\
j_{z}&=&-\partial _{t}P_{z}+\left( \vec{\nabla}_{\Phi _{0}}\times \vec{M}%
\right) _{z}+j_{(z)}{}_{conv.}
\end{eqnarray}
where the components of the convective part of the current are denoted by
the subindex $^{\prime \prime }$conv.$^{\prime \prime }$ and are derived
from the term $\frac{i}{4\mu }g^{\mu \lambda }\overline{\Psi }%
\stackrel{\leftrightarrow}{\partial}_{\lambda }$ $\Psi $. The operator $\vec{\nabla%
}_{\Phi _{0}}$ 
\begin{equation}
\vec{\nabla}_{\Phi _{0}}\equiv \ \widehat{e}_{\rho }\frac{\partial }{%
\partial \rho }+\frac{1+2\Phi _{0}}{\rho }\widehat{e}_{\varphi }\frac{%
\partial }{\partial \varphi }+(1-\Phi _{0})\widehat{e}_{z}\frac{\partial }{%
\partial z}
\end{equation}%
is the gradient operator in the gauge vacuumless string. The components of
the polarization\ and magnetization densities are given by 
\begin{eqnarray}
P_{\rho } &=&\frac{i}{2\mu }\overline{\Psi }\gamma _{(0)}\gamma _{(\rho
)}\Psi ,  \label{qq} \\
P_{\varphi } &=&\frac{i}{2\mu }\overline{\Psi }\gamma _{(0)}\gamma
_{(\varphi )}\Psi ,  \label{5.53} \\
P_{z} &=&\frac{i}{2\mu }(1+\Phi _{0})\overline{\Psi }\gamma _{(0)}\gamma
_{(z)}\Psi ,  \label{vv}
\end{eqnarray}
and
\begin{eqnarray}
M_{\rho } &=&\frac{i}{4\mu }(1+\Phi _{0})\overline{\Psi }\left[ \gamma
_{(\varphi )},\gamma _{(z)}\right] \Psi ,  \label{q1} \\
M_{\varphi } &=&\frac{i}{4\mu }(1+\Phi _{0})\overline{\Psi }\left[ \gamma
_{(z )},\gamma _{(\rho)}\right] \Psi ,  \label{22} \\
M_{z} &=&\frac{i}{4\mu }(1+\Phi _{0})\overline{\Psi }\left[ \gamma _{(\rho
)},\gamma _{(\varphi )}\right] \Psi .  \label{q3}
\end{eqnarray}

\bigskip

In the presence of an external electromagnetic field, $\vec{M}$ represents
the density of the magnetization current.

The obtained results indicate us the dependence of the current on the
Newtonian potential, and therefore, with the presence of the gauge
vacuumless string.

\section{Scalar and spinor particles in the spacetime of a global vacuumless
string}

\hskip0.7cm As a second example of a string, let us consider the global
vacuumless string whose metric is written as 
\begin{equation}
ds^{2}=\left( 1+2\Phi \right) \left( -dt^{2}+dz^{2}\right) +d\rho
^{2}+\left( 1+K\Phi \right) \rho ^{2}d\varphi ^{2},  \label{5.15.1}
\end{equation}
where $K\sim 1$ is a numerical coefficient and $\Phi (\rho )$ is the
Newtonian potential given by $\Phi (\rho )=-\kappa M^{2}\left( \frac{\rho }{%
\delta }\right) ^{\frac{4}{n+2}}$with $\kappa $ being a coefficient equal to 
$15.5$, for $n=2.$

As in the case of a gauge vacuumless string, the solution of the
Klein-Gordon equation, $\Psi (t,\rho ,\varphi ,z),$ can be separated as
\begin{equation}
\ \Psi (t,\rho ,\varphi ,z)=R(\rho )e^{-iEt+im\varphi +ikz},
\end{equation}
where $R(\rho )$ \ obeys the following equation

\begin{equation}
\frac{d^{2}R}{d\rho ^{2}}+\frac{1}{\rho }\frac{dR}{d\rho }-\left[ \frac{m^{2}%
}{\rho ^{2}}\left( 1-K\Phi \right) +\left( 1-2\Phi \right) \left(
E^{2}-k^{2}\right) -\mu ^{2}\right] R=0,  \label{5.18}
\end{equation}%
whose solution is

\begin{equation}
R(\rho )=C_{1}J_{\left\vert \nu \right\vert }(\widetilde{k}\rho
)+C_{2}N_{\left\vert \nu \right\vert }(\widetilde{k}\rho ),  \label{19}
\end{equation}%
where $\nu =\sqrt{\left( 1+K\Phi _{0}\right) m^{2}}$ \ \ \ and we have
assumed that \ the particle is restricted to move in a cylindrical narrow
region in which the Newtonian potential, $\Phi _{0},$ is approximately
constant.

In this case, the energy spectrum is given by

\begin{equation}
E=\sqrt{k^{2}+\mu ^{2}(1+2\Phi _{0})+\frac{2\left( 2+K\Phi _{0}\right) m^{2}%
}{\rho _{1}^{2}}-\frac{(1+2\Phi _{0})}{4\rho _{1}^{2}}},
\end{equation}
which shows up the dependence on the Newtonian potential $\Phi _{0}.$

Now, we will consider a spinor particle in the spacetime generated by a
global vacuumless string. To solve the Dirac equation in this background, we
will choose the following tetrads

\ 
\[
\begin{array}{ll}
e_{(0)}^{\mu } & =\left( 1+\Phi _{0}\right) \delta _{(0)}^{\mu }\vspace{0.3cm%
} \\ 
e_{(1)}^{\mu } & =\cos \varphi \delta _{(1)}^{\mu }-\frac{\left( 1-\frac{K}{2%
}\Phi _{0}\right) }{\rho }\sin \varphi \delta _{(2)}^{\mu }\vspace{0.3cm} \\ 
e_{(2)}^{\mu } & =\sin \varphi \delta _{(1)}^{\mu }+\frac{\left( 1-\frac{K}{2%
}\Phi _{0}\right) }{\rho }\cos \varphi \delta _{(2)}^{\mu }\vspace{0.3cm} \\ 
e_{(3)}^{\mu } & =\left( 1+\Phi _{0}\right) \delta _{(3)}^{\mu }%
\end{array}%
\]

Thus, the Dirac equation in this background is given by
\begin{eqnarray}
&&i\left[ \left( 1-\Phi _{0}\right) \gamma ^{(0)}\frac{\partial }{\partial t}%
+\gamma ^{(\rho )}\left( \frac{\partial }{\partial \rho }-\frac{\Phi _{0}}{%
\rho }\right) \right.  \nonumber \\
&&\qquad \qquad \left. +\frac{1-\frac{K}{2}\Phi _{0}}{\rho }\gamma
^{(\varphi )}\frac{\partial }{\partial \varphi }+\left( 1-\Phi _{0}\right)
\gamma ^{(z)}\frac{\partial }{\partial z}\right] \Psi -\mu \Psi =0,
\label{5.39}
\end{eqnarray}%
where we have assumed, once again, that the particle is restricted to move
in a narrow cylindrical region, around the string, of radii $\rho _{1}>0$ \
\ and $\rho _{2}$, with $\rho _{1}\cong $ $\rho _{2}>\rho _{1}$.

\bigskip Choosing the following Ansatz 
\begin{equation}
\Psi =\left[ 
\begin{array}{l}
\left( \sqrt{E+\mu }\right) u_{1}(\rho ) \\ 
\left( i\sqrt{E-\mu }\right) u_{2}(\rho )%
\end{array}%
\right] e^{-iEt+im\varphi +ikz},
\end{equation}%
considering the same representation of the Dirac matrices of the previous
section and taking into account that the present problem can be treated as
bi-dimensional, we find the following equations for $u_{1}(\rho )$ and $%
u_{2}(\rho )$ 
\begin{equation}
\left\{ \frac{d^{2}}{d\rho ^{2}}+\frac{1}{\rho }\frac{d}{d\rho }-\frac{1}{%
\rho ^{2}}\left[ \left( m+\frac{1}{2}\right) (1+\frac{K}{2}\Phi _{0})-\frac{1%
}{2}\right] ^{2}+(E^{2}-\mu ^{2})\right\} u_{1}(\rho )=0  \label{5.42}
\end{equation}%
and 
\begin{equation}
\left\{ \frac{d^{2}}{d\rho ^{2}}+\frac{1}{\rho }\frac{d}{d\rho }-\frac{1}{%
\rho ^{2}}\left[ \left( m+\frac{1}{2}\right) (1+\frac{K}{2}\Phi _{0})+\frac{1%
}{2}\right] ^{2}+(E^{2}-\mu ^{2})\right\} u_{2}(\rho )=0.
\label{5.43}
\end{equation}

Solutions of eqs.(\ref{5.42}) and (\ref{5.43}) are given in the general form
by 
\begin{equation}
u_{i}(\rho )=C_{i,m}^{(1)}J_{\left\vert v+(i-1)\right\vert }(\varsigma \rho
)+C_{i,m}^{(2)}N_{\left\vert v+(i-1)\right\vert }(\varsigma \rho ),
\end{equation}%
where $i=1,2$ , $\varsigma$ was defined in the previous section and $v_{i}=\left( m+\frac{1}{2}%
\right) (1+\frac{K}{2}\Phi _{0})-\frac{1}{2}$ , $C_{i,m}^{(1)}$ and $%
C_{i,m}^{(2)}$ are constant bi-spinors.

The current can be calculated following the same steps of the previous
section and taking into account the appropriate modifications. Doing this we
get the following results for the components of the current
\begin{eqnarray}
j_{t}&=&\vec{\nabla}_{\Phi _{0}}.\vec{P}+j_{(t)}{}_{conv.}\\
j_{\rho }&=&-\partial _{t}P_{\rho }+\left( \vec{\nabla}_{\Phi _{0}}\times \vec{%
M}\right) _{\rho }+j_{(\rho )}{}_{conv.}\\
j_{\varphi }&=&-\partial _{t}P_{\varphi }+\left( \vec{\nabla}_{\Phi
_{0}}\times \vec{M}\right) _{\varphi }+j_{(\varphi )}{}_{conv.}\\
j_{z}&=&-\partial _{t}P_{z}+\left( \vec{\nabla}_{\Phi _{0}}\times \vec{M}%
\right) _{z}+j_{(z)}{}_{conv.}
\end{eqnarray}

\bigskip

The operator $\vec{\nabla}_{\Phi _{0}}$ is given by 
\begin{equation}
\vec{\nabla}_{\Phi _{0}}\equiv \ \widehat{e}_{\rho }\frac{\partial }{%
\partial \rho }+\frac{1-\frac{K}{2}\Phi _{0}}{\rho }\widehat{e}_{\varphi }%
\frac{\partial }{\partial \varphi }+(1-\Phi _{0})\widehat{e}_{z}\frac{%
\partial }{\partial z}
\end{equation}%
is the gradient operator in the global vacuumless string. The components of
the polarization and magnetization densities are the same as given by Eqs. (%
\ref{qq})-(\ref{vv}) and (\ref{q1})-(\ref{q3}), respectively.

The obtained results indicate us the dependence of the current on the
Newtonian potential, and therefore, with the presence of the global
vacuumless string.

\section{Scalar and spinor particles in the spacetime of a vacuumless global
monopole}

\hskip0.7cm Now, let us consider the solution that corresponds to a
vacuumless{\it \ }global monopole, which is described by a static and
spherically symmetric metric given by\cite{Cho}%
\begin{equation}
ds^{2}=-\left( 1+2\Phi \right) dt^{2}+\left( 1+K\Phi \right)
dr^{2}+r^{2}d\Omega ^{2},  \label{5.15}
\end{equation}%
where $K$ is a coefficient which depends upon the value of $n$ and it is
equal to $9.4$, for $n=2$, in the case of a vacuumless global monopole. The
Newtonian potential, $\Phi (r)$, is given by 
\[
\Phi (r)=-\kappa M^{2}\left( \frac{r}{\delta }\right) ^{\frac{4}{n+2}}, 
\]%
with $\kappa $ being a numerical coefficient which is approximately equal to
1.

Now, we will obtain the solution of the Klein-Gordon equation for a particle
in this spacetime, which can be written as
\begin{eqnarray}
&&\left\{ -\left( 1-2\Phi \right) \frac{\partial ^{2}}{\partial t^{2}}+\frac{%
1}{r^{2}\sin ^{2}\theta }\frac{\partial ^{2}}{\partial \varphi ^{2}}\right.
\bigskip  \nonumber \\
&&\quad +\left[ \frac{(K+2)}{2}\left( 1-2\left( K+1\right) \Phi \right) 
\frac{\partial \Phi }{\partial r}+\frac{2}{r}\left( 1-K\Phi \right) \right] 
\frac{\partial }{\partial r}\bigskip  \nonumber \\
&&\left. \quad -K\frac{\partial \Phi }{\partial r}\frac{\partial }{\partial r%
}+\left( 1-K\Phi \right) \frac{\partial ^{2}}{\partial r^{2}}+\frac{\cos
\theta }{r^{2}\sin \theta }\frac{\partial }{\partial \theta }+\frac{1}{r^{2}}%
\frac{\partial ^{2}}{\partial \theta ^{2}}-\mu ^{2}\right\} \Psi =0.
\label{5.26}
\end{eqnarray}

The solution of this equation can be separated as 
\begin{equation}
\Psi (t,r,\theta ,\varphi )=\frac{R(r)}{\sqrt{r}}Y_{lm}(\theta ,\varphi )
\label{5.27}
\end{equation}%
where $Y_{lm}(\theta ,\varphi )$ are the spherical harmonics.

Substituting (\ref{5.27}) into (\ref{5.26}), we obtain the following radial
equation 
\begin{eqnarray}
&&\frac{d^{2}R}{dr^{2}}+\frac{1}{r}\frac{dR}{dr}+\left\{ E^{2}\left[
1-(K+2)\Phi _{0}\right] -\mu ^{2}\left( 1+K\Phi _{0}\right) \right. \bigskip
\nonumber \\
&&\left. \qquad \qquad \qquad \qquad \qquad \qquad -\frac{l(l+1)}{r^{2}}%
\left( 1+K\Phi _{0}\right) \right\} R(r)\left. =0\right. ,  \label{jjjj}
\end{eqnarray}%
whose solution is given by in terms of Bessel functions as 
\begin{equation}
R(r)=C_{1}J_{\frac{1}{2}\sqrt{1+4v^{2}}}(\widetilde{k}r)+C_{2}N_{\frac{1}{2}%
\sqrt{1+4v^{2}}}(\widetilde{k}r)  \label{mmmm}
\end{equation}%
where $\widetilde{k}^{2}\equiv -\mu ^{2}(1-K\Phi _{0})+E^{2}\left[
1+(2+K)\Phi _{0}\right] $, $v^{2}=l(l+1)(1+K\Phi _{0})$ and $\Phi _{0}$ a
constant value for the Newtonian potential. Note that we are assuming that
the particle is restricted to move in a spherical region of radii $r_{1}>0$
and $r_{2}$ and such that $r_{1}\cong r_{2}>r_{1}$.

Analogously to the case of the vacuumless strings, we can determine the
energy spectrum which is given by 
\begin{equation}
E=\sqrt{\mu ^{2}(1+K\Phi _{0})+\frac{l(l+1)}{r_{1}^{2}}\left( 1-2\Phi
_{0}\right) +\frac{3}{4r_{1}^{2}}\left( 1-(2+K)\Phi _{0}\right) }.
\label{5.31}
\end{equation}

This result shows explicitly the dependence of the energy on the presence of
the vacuumless global monopole.

Now, we will obtain the solution of the Dirac equation in the vacuumless
global monopole spacetime and determine the current associated with the
particle of mass $\mu $. In order to solve the Dirac equation in this
spacetime we will choose the following tetrads\cite{Hai}
\begin{equation}
e_{(a)}^{\mu }=\left( 
\begin{array}{llll}
1 & 0 & 0 & 0 \\ 
0 & \left( 1-\frac{K}{2}\Phi _{0}\right) \sin \theta \cos \varphi & \frac{1}{%
r}\cos \theta \cos \varphi & -\frac{1}{r}\frac{\sin \varphi }{\sin \theta }
\\ 
0 & \left( 1-\frac{K}{2}\Phi _{0}\right) \sin \theta \sin \varphi & \frac{1}{%
r}\cos \theta \sin \varphi & \frac{1}{r}\frac{\cos \varphi }{\sin \theta }
\\ 
0 & \left( 1-\frac{K}{2}\Phi _{0}\right) \cos \theta & -\frac{1}{r}\sin
\theta & 0%
\end{array}%
\right) ,
\end{equation}%
where we also assume that the particle is restricted to move inside a very
narrow spherical region.

Thus, the Dirac equation turns into
\begin{eqnarray}
&&\left\{ \left( 1-\Phi _{0}\right) \gamma ^{0}\frac{\partial }{\partial t}%
+\left( 1-\frac{K}{2}\Phi _{0}\right) \gamma ^{r}\frac{\partial }{\partial r}%
+\frac{i}{r}\gamma ^{\theta }\frac{\partial }{\partial \theta }+\frac{%
i\gamma ^{\varphi }}{r\sin \theta }\frac{\partial }{\partial \varphi }\right.
\nonumber \\
&&\left. \hspace{6.7cm}-\frac{iK}{2r}\Phi _{0}\gamma ^{r}-\mu \right\} \Psi
=0,  \label{5.63a}
\end{eqnarray}%
where 
\begin{eqnarray}
\gamma ^{(r)} &=&\sin \theta \cos \varphi \gamma ^{(1)}+\sin \theta \sin
\varphi \gamma ^{(2)},  \nonumber \\
\gamma ^{(\theta )} &=&\cos \theta \cos \varphi \gamma ^{(1)}+\cos \theta
\sin \varphi \gamma ^{(2)}-\sin \theta \gamma ^{(3)},  \nonumber \\
\gamma ^{(\varphi )} &=&\cos \theta \gamma ^{(2)}-\sin \varphi \gamma ^{(1)}.
\end{eqnarray}

In this case, it is convenient to use the following representation of the
Dirac matrices in Minkowski spacetime 
\begin{equation}
\gamma ^{(0)}=\left( 
\begin{array}{ll}
1 & 0 \\ 
0 & -1%
\end{array}%
\right);\qquad \gamma ^{(i)}=\left( 
\begin{array}{ll}
0 & \sigma ^{i} \\ 
-\sigma ^{i} & 0%
\end{array}%
\right).
\end{equation}

The set of solutions of eq.(\ref{5.63a}) can be written in the form 
\begin{equation}
\Psi =e^{-iEt}\left[ 
\begin{array}{l}
\psi _{j,m}^{(1)} \\ 
\psi _{j,m}^{(2)}%
\end{array}%
\right]  \label{5.66}
\end{equation}%
where $\psi _{j,m}^{(1)}$ and $\psi _{j,m}^{(2)}$ are bi-spinors 
\begin{equation}
\Psi _{j,m}^{(1)}=\left[ 
\begin{array}{l}
f(r)\psi _{j,m}^{(1)} \\ 
g(r)\psi _{j,m}^{(2)}%
\end{array}%
\right]  \label{5.67}
\end{equation}%
and 
\begin{equation}
\Psi _{j,m}^{(2)}=\left[ 
\begin{array}{l}
f(r)\psi _{j,m}^{(2)} \\ 
g(r)\psi _{j,m}^{(1)}%
\end{array}%
\right]  \label{5.68}
\end{equation}%
with $\psi _{j,m}^{(1)}$ $\psi _{j,m}^{(2)}$ being spinor spherical
harmonics.

Substituting (\ref{5.67}) into the Dirac equation given by (\ref{5.63a}) ,
we obtain 
\begin{equation}
(E-\mu )f(r)=i\left[ \frac{d}{dr}+\frac{1+\left( j-\frac{1}{2}\right) \left(
1-\frac{K}{2}\Phi _{0}\right) }{r}\right] g  \label{5.69}
\end{equation}%
and 
\begin{equation}
(E+\mu )g(r)=i\left[ \frac{d}{dr}+\frac{1-\left( j+\frac{1}{2}\right) \left(
1-\frac{K}{2}\Phi _{0}\right) }{r}\right] f.  \label{5.70}
\end{equation}%
Combining eqs. (\ref{5.69}) and (\ref{5.70}) , we find that the solutions of
the resulting equations are given by 
\begin{equation}
\hspace{1.3cm}f(r)=\frac{C_{1}}{\sqrt{r}}J_{v-\frac{1}{2}}(\varsigma r)+\frac{C_{2}}{%
\sqrt{r}}N_{v-\frac{1}{2}}(\varsigma r)
\end{equation}%
and 
\begin{equation}
g(r)=\frac{C_{1}}{\sqrt{r}}J_{v+\frac{1}{2}}(\varsigma r)+\frac{C_{2}}{\sqrt{r}}N_{v+%
\frac{1}{2}}(\varsigma r),
\end{equation}%
where $v=\left( j+\frac{1}{2}\right) \left( 1+%
\frac{K}{2}\Phi _{0}\right) $ .

Proceeding in analogy with the previous case, we can determine the current.
Thus, we have
\begin{equation}
\begin{array}{l}
\begin{array}{ll}
\left[ \gamma ^{\mu },\partial _{\lambda }\gamma ^{3}\right]  & =-\frac{2}{%
r^{2}\sin \theta }\left( 1-\frac{K\Phi _{0}}{2}\right) \gamma ^{(1)}\gamma
^{(3)}-\frac{2}{r^{2}\sin ^{2}\theta }\gamma ^{(2)}\gamma ^{(3)},%
\end{array}%
\bigskip  \\ 
\begin{array}{ll}
\left[ \gamma ^{\mu },\partial _{\lambda }\gamma ^{2}\right]  & =-\frac{1}{%
r^{2}}\left( 1-\frac{K\Phi _{0}}{2}\right) \gamma ^{(1)}\gamma ^{(2)},%
\end{array}%
\bigskip  \\ 
\begin{array}{ll}
\left[ \gamma ^{\lambda }\Gamma _{\lambda },\gamma ^{3}\right]  & =\frac{1}{%
r\sin \theta }\gamma ^{(1)}\gamma ^{(3)}+\frac{\cot \theta }{r^{2}\sin
\theta }\gamma ^{(2)}\gamma ^{(3)},%
\end{array}%
\bigskip  \\ 
\begin{array}{ll}
\left[ \gamma ^{\lambda }\Gamma _{\lambda },\gamma ^{2}\right]  & =\frac{1}{%
4r^{2}}\gamma ^{(1)}\gamma ^{(2)}+\frac{\cot \theta }{r^{2}}\gamma
^{(2)}\gamma ^{(2)},%
\end{array}%
\bigskip  \\ 
\begin{array}{ll}
\left[ \gamma ^{\lambda }\Gamma _{\lambda },\gamma ^{1}\right]  & =\frac{4}{r%
}\left( 1-\frac{K\Phi _{0}}{2}\right) \gamma ^{(1)}\gamma ^{(1)}+\left( 1-%
\frac{K\Phi _{0}}{2}\right) \frac{\cot \theta }{r}\gamma ^{(2)}\gamma
^{(1)},%
\end{array}%
\bigskip  \\ 
\begin{array}{ll}
\left[ \gamma ^{\lambda }\Gamma _{\lambda },\gamma ^{0}\right]  & =\frac{4}{r%
}\gamma ^{(1)}\gamma ^{(0)}+\frac{\cot \theta }{r}\gamma ^{(2)}\gamma
^{(0)},%
\end{array}%
\end{array}
\label{5.76}
\end{equation}%
and 
\begin{equation}
\begin{array}{ll}
\sigma ^{01} & =i\left( 1-\frac{K}{2}\Phi _{0}\right) \gamma ^{(0)}\gamma
^{(1)},\bigskip  \\ 
\sigma ^{02} & =\frac{i}{r}(1+\Phi _{0})\gamma ^{(0)}\gamma ^{(2)},\bigskip  \\ 
\sigma ^{03} & =\frac{i}{r\sin \theta }(1-\Phi _{0})\gamma ^{(0)}\gamma
^{(3)},\bigskip  \\ 
\sigma ^{12} & =\frac{i}{r}\left( 1-\frac{K}{2}\Phi _{0}\right) \gamma
^{(1)}\gamma ^{(2)},\bigskip  \\ 
\sigma ^{13} & =\frac{i}{r\sin \theta }\left( 1-\frac{K}{2}\Phi _{0}\right)
\gamma ^{(1)}\gamma ^{(3)},\bigskip  \\ 
\sigma ^{23} & =\frac{i}{r\sin \theta }(1-\Phi _{0})\gamma ^{(2)}\gamma
^{(3)}.%
\end{array}
\label{5.77}
\end{equation}%
Therefore, using the expression for the current given by the eq.(\ref{5.46})
and the results given in (\ref{5.76}) and (\ref{5.77}), we obtain the
following expressions for the components of the current

\begin{eqnarray}
j_{t} &=&\vec{\nabla}_{\Phi _{0}}.\vec{P}+j_{(t)}\rho _{conv.}  \label{5.78}
\\
j_{r} &=&-\partial _{t}P_{r}+\left( \vec{\nabla}_{\Phi _{0}}\times \vec{M}%
\right) _{r}+j_{(r)}{}_{conv.}  \label{5.79} \\
j_{\theta } &=&-\partial _{t}P_{\theta }+\left( \vec{\nabla}_{\Phi
_{0}}\times \vec{M}\right) _{\theta }+j_{(\theta )}{}_{conv.}  \label{5.80}
\\
j_{z} &=&-\partial _{t}P_{z}+\left( \vec{\nabla}_{\Phi _{0}}\times \vec{M}%
\right) _{z}+j_{(z)}{}_{conv.}  \label{5.81}
\end{eqnarray}%
where 
\begin{equation}
\vec{\nabla}_{\Phi _{0}}\equiv \ \left( 1-\frac{K}{2}\Phi _{0}\right) 
\widehat{e}_{r}\frac{\partial }{\partial r}+\frac{1}{r}\widehat{e}_{\theta }%
\frac{\partial }{\partial \theta }+\frac{1}{r\sin \theta }\widehat{e}%
_{\varphi }\frac{\partial }{\partial \varphi }
\end{equation}%
is the gradient operator in the vacuumless global monopole spacetime. The
components of the polarization and magnetization densities are given by

\begin{eqnarray}
P_{r} &=&\frac{i}{2\mu }\overline{\Psi }\gamma _{(0)}\gamma _{(r)}\Psi
\bigskip ,  \nonumber \\
P_{\theta } &=&\frac{i}{2\mu }\overline{\Psi }\gamma _{(0)}\gamma _{(\theta
)}\Psi \bigskip ,  \label{5.82} \\
P_{\varphi } &=&\frac{i}{2\mu }\overline{\Psi }\gamma _{(0)}\gamma
_{(\varphi )}\Psi ,  \nonumber
\end{eqnarray}
and
\begin{eqnarray}
M_{r} &=&\frac{i}{4\mu }\overline{\Psi }\left[ \gamma _{(\varphi )},\gamma
_{(z)}\right] \Psi \bigskip ,  \nonumber \\
M_{\theta } &=&\frac{i}{4\mu }\overline{\Psi }\left[ \gamma _{(z)},\gamma
_{(r)}\right] \Psi \bigskip ,  \label{5.83} \\
M_{z} &=&\frac{i}{4\mu }\overline{\Psi }\left[ \gamma _{(r)},\gamma
_{(\varphi )}\right] \Psi .  \nonumber
\end{eqnarray}

Note that the expressions for the components of the current given by eq.(\ref%
{5.78}), (\ref{5.79}), (\ref{5.80}) and (\ref{5.81}) shows explicitly the
influenced of the gravitational field of the vacuumless global monopole on
this physical quantity.

\section{Conclusions}

\hskip0.7cm In this paper we solved Klein-Gordon and Dirac equations in the
gauge and global vacuumless strings and in the vacuumless global monopole
spacetimes, and showed that the solutions of these equations depend on the
Newtonian potentials associated each one of these spacetimes. We also
calculated the energy spectra and \ the currents, for the scalar and spinor
cases, respectively, and showed their dependence on the respective Newtonian
potential. In the obtained results we have made the hypotheses that the
particles are restricted to move in very narrow regions, so that the
Newtonian potentials are approximately constant. These hypotheses are
reasonable, from the physical point of view, since we are treating with
gravitational fields in the weak field approximation.

\bigskip

{\bf {Acknowledgments}} \newline
\newline
We would like to thank Conselho Nacional de Desenvolvimento Cient\'{\i}fico
and Tecnol\'{o}gico (CNPq) and Funda\c{c}\~{a}o de Apoio \`{a} Pesquisa do
Estado da Para\'{\i}ba (FAPESQ) /CNPq (PRONEX) for partial financial support.

\end{document}